\documentclass[12pt]{article}
\usepackage[dvips]{epsfig}

\begin{document}
\setlength{\unitlength}{1mm}

\newcommand{\ba} {\begin{eqnarray}}
\newcommand{\ea} {\end{eqnarray}}
\newcommand{\be}{\begin{equation}}
\newcommand{\ee}{\end{equation}}
\newcommand{\n}[1]{\label{#1}}
\newcommand{\eq}[1]{Eq.(\ref{#1})}
\newcommand{\ind}[1]{\mbox{\tiny{#1}}}
\renewcommand\theequation{\thesection.\arabic{equation}}

\newcommand{\nn}{\nonumber \\ \nonumber \\}
\newcommand{\nl}{\\  \nonumber \\}
\newcommand{\pr}{\partial}
\renewcommand{\vec}[1]{\mbox{\boldmath$#1$}}

\author{Claude Barrab\`es\thanks{E-mail : barrabes@phys.univ-tours.fr}\\     
\small Laboratoire de Math\'ematiques et Physique Th\'eorique\\
\small  CNRS/UMR 6083, Universit\'e F. Rabelais, 37200 Tours, France\\
Werner Israel\thanks{E-mail : israel@uvic.ca}\\
\small Department of Physics and Astronomy\\
\small  University of Victoria, Canada, V8W 3P6}
\title{Lagrangian Brane dynamics in general relativity 
and Einstein-Gauss-Bonnet gravity}
\date{}
\maketitle

\begin{abstract}

This paper gives a new, simple and concise derivation of brane actions and
brane dynamics in general relativity and in Einstein-Gauss-Bonnet gravity.
We present a unified treatment, applicable to timelike surface layers and
spacelike transition layers, and including consideration of the more 
difficult lightlike case.

\end{abstract}
\thispagestyle{empty}

\section{Introduction}
\setcounter{equation}0
Thin walls and shells of matter have surfaced increasingly in a variety of
situations in astrophysics, cosmology and quantum gravity. 
Highly compressed expanding shells of material emerge from supernova 
explosions \cite{Pim}, as false-vacuum bubble walls in inflationary
phase transitions \cite{Col}, and in hypothetical scenarios of 
``new-universe'' creation \cite{BF}. In a sudden global phase transition, the
transition region can sometimes be idealized as an infinitesimaly 
thin spacelike surface layer \cite{FMM}.
Theoretical exploration of basic issues 
of principle, such as the possible outcomes of a classical or quantum 
gravitational collapse \cite{Haj}, are often simplified, for purposes 
of a first reconnaissance, by idealizing the collapsing matter as a 
thin shell, thus reducing the complex differential field equations 
to simple algebraic junction conditions.

With the advent of brane-world scenarios \cite{Maa}, the scope of 
thin-shell dynamics has broadened to embrace higher dimensions and 
string-inspired extensions of Einsteinian gravity, in particular 
Einstein-Gauss-Bonnet (EGB) theory \cite{Zw}. Recently, a class
of Weyl-conformally invariant $p$-brane theories, which includes
lightlike branes, has been proposed \cite{Guen}.

For purely Einsteinian shells the classical dynamics is straightforward 
\cite{MTW} (though the variational and Hamiltonian aspects can be subtle
\cite{Muk}). Appending Gauss-Bonnet terms to the usual Einstein-Hilbert
action, however, is attended by a considerable step-up in complexity.
Initially, there were even doubts whether EGB shells admit a 
distributional description at all: the EGB bulk field equations 
develop ill-defined products of delta- and step-functions 
in the thin-layer limit unless
the terms are arranged with care \cite{Der}. The key is to express the
field equations in canonical form, with distance from the layer in the 
role of ``time''; this segregates the most singular terms into the 
canonical momentum. The canonical momentum is the dynamical variable that 
``jumps'' cleanly at a thin layer.

The key role of canonical momentum suggets that the EGB junction 
conditions are most easily derived from the action. This derivation 
was carried through by Davis \cite{Dav}, so that the basic equations of 
EGB shell dynamics are now well-established and widely employed \cite{DerS}. 
Still to be desired is a systematic, self-contained exposition which
draws together general-relativistic and EGB shell dynamics within a 
unified Lagrangian framework, and includes consideration of the lightlike
limit \cite{BIH}. We hope this paper will go some way toward filling this gap.

\section{Toy Model}
\setcounter{equation}0
To illustrate the essential ideas, we take a simple example from 
one-dimensional particle mechanics. We choose a ``bulk'' action functional
of the path $q=q(t)$,
\be
S_{bulk}[q]\,=\,\int^{t_f}_{t_i}\,(L-V_{ext})\,dt\, ,\n{2.1}
\ee
with an acceleration-dependent Lagrangian of the form
\be
L\,=\,-\,b(q)\,\ddot{q}\,-\frac{1}{2}\,b'(q)\,\dot{q}^2\,-V(q)\, ,\n{2.2}
\ee
where the functions $b\,,V$ are arbitrary and $V_{ext}(q,t)$ is an 
arbitrary external potential.

We are considering $q$ as an analogue of the metric, $t$ as an analogue
of distance normal to a boundary surface or layer, and $\dot{q}$ as 
an analogue of extrinsic curvature $K\sim \partial g/\partial n $.
The two terms of (\ref{2.1}) simulate the geometrical and matter actions;
the external force $F=-\partial V_{ext}/\partial q $ is the analogue of
material stress-energy $T_{\alpha\beta}$. The particular, quasilinear
functional form (\ref{2.2}) is patterned after the Einstein-Hilbert (EH)
and Gauss-Bonnet (GB) Lagrangians. (The EH Lagrangian is quasilinear 
in the narrow sense that the coefficients of the second-derivative
terms are functions of the metric only, not its first derivatives. 
But the corresponding coefficients do depend on first derivatives in 
the case of the GB Lagrangian).

Although $L$ involves second derivatives, its quasilinearity ensures 
that the Euler-Lagrange equation
\be
\frac{\not\hspace{-0.1em}\partial L}{\not\hspace{-0.1em}\partial q}\,
+\,F\,=0\, ,\n{2.3}
\ee
for the classical path is no higher than second order. The classical path
is thus uniquely determined by fixing its two endpoints $q_i $, $q_f $. 
But this path does not extremize the bulk action (\ref{2.1}), because the
endpoint velocities  $\dot{q}_i $, $\dot{q}_f $ can still be varied
freely: we have
\be
\delta S_{bulk}[q]\,=\, \int^{t_f}_{t_i}\,
\left( \frac{\not\hspace{-0.1em}\partial L}{\not\hspace{-0.1em}\partial q}\,
+F \right)\,\delta q(t)\,dt\,\,+
\left[ \,p\,\delta q\,-\delta B(q,\dot{q})\, \right]^{t_f}_{t_i}\, ,\n{2.4}
\ee
where
\be
p\,=\,b'(q)\,\dot{q}\,\,;\qquad B(q,\dot{q})\,=\,b(q)\,\dot{q}\, .\n{2.5}
\ee   

To have an action that is extremized by the classical path, one must add 
a boundary term \cite{Reg} to the bulk action (\ref{2.1}):
\be
S[q]\,=\,S_{bulk}[q]\,+\,\left. B(q,\dot{q})\,\right|^{t_f}_{t_i}\, ,\n{2.6}
\ee
The extremal of the action (\ref{2.6}) now depends solely on the 
endpoints $q_i $, $q_f $ of the classical path, in accordance with the 
Hamilton-Jacobi equation
\be
\delta S_{extrem}(q_i ,t_i ;q_f ,t_f )\,=\,p\,
\left.\delta q\,\right|^{t_f}_{t_i}\, ,\n{2.7}
\ee

More fundamentally, the boundary term $B$ is needed to preserve 
the composition law
\be
S[1 \to 2 \to 3]\,=\,S[1 \to 2]\,+\,S[2 \to 3]\,, \n{2.8}
\ee
for an arbitrary continuous path joining the points $1\,,2\,,3$ 
with a sharp bend at $2$ (as used for example in the ``zig-zag'' definition
of the path integral \cite{Fey}).

The Euler-Lagrange equation (\ref{2.3}) can be expressed as
\be
\frac{dp}{dt}\,+\frac{\partial V}{\partial q}\,=\,F(q,t)\,.\n{2.9}
\ee
This is the analogue of the gravitational field equations, with the 
right-hand side representing the stress-energy of matter. 
To simulate a thin surface layer, we consider an impulsive force 
acting at time $t_0$:
\be
F(q,t)\,=\,\sigma (q)\,\delta (t-t_0)\,.\n{2.10}
\ee
This will produce a discontinuity in the momentum $p$. Since all
delta-function contributions to the left-hand side are gathered into
$dp/dt$, it is straightforward to integrate the equation of motion
(\ref{2.9}) to obtain the jump across the discontinuity:
\be
\lbrack\hspace{-0.1em}\vert\,p\,\vert\hspace{-0.1em}\rbrack\,
=\,\sigma (q)\,,\n{2.11}
\ee
where $\lbrack\hspace{-0.1em}\vert\,p\,\vert\hspace{-0.1em}\rbrack\,=\,
\lim_{\epsilon \to 0} \{ p(t_0 +\epsilon )-p(t_0 -\epsilon )\} $. 
This is the analogue of the geometrical junction conditions at a 
surface layer. The key to deriving it is just the identification 
of the canonical momentum $p$ -- most easily from the Hamilton-Jacobi 
equation (\ref{2.7}), or from a suitable adaptation of Lagrange's 
definition applied to an equivalent first-order Lagrangian, 
see (\ref{2.16}) below.

The jump condition(\ref{2.11}) would follow as
\be
\frac{\partial S_{imp} (q_0 ,\dot{q}_0^{\pm} )}{\partial q_0 }\,=\,0\,,\n{2.12}
\ee
from an ``impulsive action'' $S_{imp}$, as a function of position
$q(t_0 )\,=\,q_0 $ and pre- and post-shock velocities
$\dot{q}(t_0 \pm \epsilon )\,=\,\dot{q}_0^{\pm} $. This is given by
\be
S_{imp}\,=\,-\left[ \,B(q,\dot{q})\, \right]^{t_0 +\epsilon}_{t_0 -\epsilon}\,
-\,\int^{t_0 +\epsilon}_{t_0 -\epsilon}\,V_{ext}(q,t)\,dt\,,\n{2.13}
\ee
and it coincides with the bulk action (\ref{2.1}) if we choose 
$t_f\,=\,t_0 +\epsilon $, $t_i\,=\,t_0 -\epsilon $. Alternatively,
(\ref{2.13}) is obtainable without integrating $L$ through the shock,
instead considering $t_0 -\epsilon $ and $t_0 +\epsilon $ as future
 and past endpoints respectively, each with its own boundary action $B$,
so that the bulk action (\ref{2.6}) becomes
\be
S[q]\,=\,
\left( \int^{t_0 -\epsilon}_{t_i}\,+\,\int^{t_f}_{t_0 +\epsilon}\right)\,L\,dt
\,\,-\int^{t_f}_{t_i}\,V_{ext}\,dt\,\,+\,[\,B\,]^{t_0 -\epsilon}_{t_i}\,+\,
[\,B\,]^{t_f}_{t_0 +\epsilon}\,,\n{2.14}
\ee
when an impulse acts at $t_0 $.

It should be noted that the close relationship (\ref{2.13}), (\ref{2.14})
of the boundary action $B$ to an impulsive action hinges on the special form
(\ref{2.2}) of the action, and does not extend to an arbitrary quasilinear 
Lagrangian. It does, nevertheless carry over to the EH action, and also to the
GB action modulo removable $K^2\,\partial K/\partial n $ terms. The
origins of this peculiar circumstances will emerge in Sec.3.

The bulk + boundary action (\ref{2.6}) is really a thinly disguised 
first-order action:
\be
S[q]\,=\,\int^{t_f}_{t_i}\,(L_1 -V_{ext})\,dt\,;\qquad
L_1\,=\,\frac{1}{2}\,b'(q)\,\dot{q}^2 \,-V(q)\,,\n{2.15}
\ee
with the standard definition
\be
p\,=\,\frac{\partial L_1 }{\partial \dot{q}}\,.\n{2.16}
\ee
Therewith everything relating to this mechanical model takes on a trivial
appearance. Not so, however, for its gravitational counterparts: there,
only the original, second-order Lagrangian is a geometrical object and 
a scalar; the split into a first-order Lagrangian and a pure divergence 
cannot be made in a co-ordinate invariant and boundary-independent
way. One is essentially forced to retain the bulk + boundary formulation.

Let us finally note that, because of the freedom to redefine the bulk 
part of the total action by adding a total derivative, the definitions of 
canonical momentum and boundary action are (trivially) arbitrary to 
the extent 
\be
L_1\,\to\,L_1\,+\,\frac{d}{dt}f(q)\,;\qquad B\,\to\,B\,-\,f(q)\,;\qquad
p\,\to\,p\,+\,f'(q)\,.\n{2.17}
\ee
This has no effect on the impulsive jump conditions (\ref{2.11}), because 
the arbitrary function $f(q)$ is continuous at the shock.

\section{Brane Dynamics: Einstein-Hilbert Action}
\setcounter{equation}0
We begin by focussing on general relativity and on higher-dimensional 
gravitational theories governed by the Einstein-Hilbert action. Our purpose
is to derive the well-known junction conditions \cite{MTW}, \cite{BIH}
which determine the motion of a surface layer in such theories from
the action. We shall present a unified treatment, applicable to timelike, 
spacelike or lightlike layers.

In an $(n+1)$-dimensional spacetime, the Einstein-Hilbert bulk action is
\be
S_{bulk}\,=\,\frac{1}{2\kappa}\,\int\,{\cal L}_{EH}(g,\Gamma)\,
d^{n+1}x\,, \n{3.1}
\ee
where
\be
{\cal L}_{EH}\,=\,\sqrt{-g}\,R\,=\,\sqrt{-g}\,g^{\mu\nu}\,
(\partial_{\lambda}\Gamma^{\lambda}_{\mu\nu}\,-\, \partial_{\nu}
\Gamma^{\alpha}_{\mu\alpha})\,-\,{\cal L}_1\,,\n{3.2}
\ee
\be
{\cal L}_1\,=\,\sqrt{-g}\,g^{\mu\nu}\,
(\Gamma^{\alpha}_{\mu\beta}\,\Gamma^{\beta}_{\nu\alpha}\,-\,
\Gamma^{\alpha}_{\mu\nu}\,\Gamma^{\beta}_{\alpha\beta}\,)\, .\n{3.3}
\ee
The second-order Lagrangian (\ref{3.2}) is degenerate, and can be reduced 
to first-order form by extracting a pure divergence:
\be
{\cal L}_{EH}\,=\,-\partial_{\lambda}\sigma^{\lambda}\,+\,{\cal L}_1\,,\n{3.4}
\ee
where
\be
\sigma^{\lambda}(g,\Gamma)\,\equiv \,\sqrt{-g}\,(g^{\lambda\mu}\,
\Gamma^{\alpha}_{\mu\alpha}\,-\,g^{\mu\nu}\,
\Gamma^{\lambda}_{\mu\nu}\,)\, ,\n{3.5}
\ee
which is reducible to
\be
\sigma^{\lambda}\,=\,\frac{1}{\sqrt{-g}}\,\partial_{\mu}(-g\,g^{\lambda\mu})\,.
\n{3.6}
\ee
The first-order Lagrangian ${\cal L}_1 $ dates back to Lorentz, Hilbert,
Einstein, Weyl and Felix Klein, and was employed by Einstein \cite{Ein} to
define his pseudo-tensor for gravitational energy and radiation.

The complete Einstein-Hilbert action $S_{EH}$ complements the bulk action 
(\ref{3.1}) with a term $S_{bdy}$ coming from the boundary, which soaks up
the pure divergence in (\ref{3.4}) \cite{Ein}. Including also the 
matter contribution, the complete action reads
\be
S_{EH}\,=\,S_{bulk}\,+\,S_{bdy}\,+\,S_{mat}\,=\,
\frac{1}{2\kappa}\,\int\,{\cal L}_1\,d^{n+1}x\,+\,\,S_{mat}\,.\n{3.7}
\ee
Here
\be
S_{bdy}[g,\Gamma ]\,=\,\frac{1}{2\kappa}\,\int\,
\partial_{\lambda}\sigma^{\lambda}\,d^{n+1}x\,,\n{3.8}
\ee
is convertible to a surface integral over the boundary $\Sigma $, and
$S_{mat}$ yields the material stress-energy tensor
\be
\frac{\delta S_{mat}}{\delta g^{\mu\nu}}\,=\,-\frac{1}{2}\,\sqrt{-g}\,
T_{\mu\nu}\,.\n{3.9}
\ee

Computing the variation of the bulk action (\ref{3.1}) is facilitated by the 
Palatini identity
\be
\delta R_{\mu\nu}\,=\,\delta \Gamma^{\lambda}_{\mu\nu |\lambda}\,-\,
\delta\Gamma^{\alpha}_{\mu\alpha |\nu}\,,\n{3.10}
\ee
where $|$ stands for the covariant derivative associated with 
the metric $g_{\mu\nu}$. 
Recalling (\ref{3.5}), and momentarily treating $g$ and $\Gamma $ as
independent, this gives
\be
\delta{\cal L}_{EH}(g,\Gamma )\,=\, (\delta _g\,+\,\delta_{\Gamma}) 
\lbrace \sqrt{-g}\,g^{\mu\nu}\,R_{\mu\nu}(\Gamma )\rbrace\,=\,
\delta g^{\mu\nu}\sqrt{-g}\,G_{\mu\nu}\,-\,
\partial_{\lambda}(\delta_{\Gamma}\sigma^{\lambda})\,.\n{3.11}
\ee
From (\ref{3.7})--(\ref{3.9}) we thus obtain
\be
\delta S_{EH}\,=\,\frac{1}{2\kappa}\,\int\,
(G_{\mu\nu}\,-\,\kappa\,T_{\mu\nu})\,
\delta g^{\mu\nu}\sqrt{-g}\,d^{n+1}x\,+\,
\delta _g \,S_{bdy}[g,\Gamma ]\,. \n{3.12}
\ee
The last term depends only on the metric variation $\delta g_{\alpha\beta}$
at the boundary, not $\delta \Gamma $. Thus, for metric variations 
which vanish at the boundary, we find
\be
\frac{\delta S_{EH}}{\delta g^{\mu\nu}}\,=\,0\qquad \Rightarrow \qquad 
G_{\mu\nu}\,=\,\kappa T_{\mu\nu}\, ,\n{3.13}
\ee
which are the bulk equations.

While the second-order Lagrangian (\ref{3.2}) is a geometrical invariant, 
its split (\ref{3.4}) into a first-order part and a pure divergence is 
co-ordinate-dependent. This split can, however, be endowed with 
geometrical significance by a special choice of the bulk
co-ordinates $x^{\mu}$ which anchors them to the
boundary $\Sigma $. In general, $\Sigma $ will be characterized by parametric
equations 
\be
\Sigma \,:\,x^{\mu}\,=\,x^{\mu}(\xi^a ,z)\,\,;\qquad 
z(x^{\mu})\,=\,0\,,\n{3.14}
\ee
where $\xi^a\,(a=1,...n)$ are arbitrary intrinsic co-ordinates. Then 
(\ref{3.8}) integrates to 
\be
S_{bdy}\,=\,\frac{1}{2\kappa}\,\int_{\Sigma}\,\sigma^{\lambda}\,
\partial_{\lambda}z\,\,dA\,,\n{3.15}
\ee
where $dA$ is a (non-invariant) element of $n$-area. We now impose 
the anchoring condition on $x^{\mu} $:
\be
\Sigma \,:\, x^0\,\equiv\,z\,=\,0\,\,,\,\,x^a\,=\,\xi^a\,.\n{3.16}
\ee
Then (\ref{3.15})
becomes
\be
S_{bdy}\,=\,\frac{1}{2\kappa}\,\int_{\Sigma}\,\sigma^0 (g,\Gamma )\,d^n \xi\,,
\n{3.17}
\ee
with
\be
\sigma^0\,=\,\sigma^{\lambda}\,\partial_{\lambda}z\,=\,
(-g)^{-\frac{1}{2}}\,\partial_{\lambda}\lbrack (-g)\,g^{\lambda\mu}\,
\partial_{\mu}z \rbrack\,.\n{3.18}
\ee

Its non-covariant appearance notwithstanding, (\ref{3.18}) is actually (twice)
the mean extrinsic curvature density of $\Sigma $, up to a dynamically 
irrelevant term. Moreover, it is regular even for a lightlike $\Sigma $, 
so that (\ref{3.18}) may be considered a valid extension of the notion of
mean extrinsic curvature density to the lightlike case. 

To verify these statements, suppose first that $\Sigma $ is non-lightlike, 
and thus has a unit normal $n^{\mu}$, which is 
transverse to $\Sigma $, say in the
direction of increasing $z$, $(n^0 >0 )$; $n.n\,=\,\epsilon $
with $\epsilon =+(-)1$ whenever $\Sigma $ is time(space)like. 
Then (\ref{3.18}) can be shown to reduce to 
\be
\sigma^0\,=\,2\,\epsilon\,{\cal K}\,+\,\sqrt{-g}\,\epsilon\,N^{-2}\,
\partial_a s^a\,\,;\qquad
{\cal K}\,=\,|{}^{(n)} g|^{\frac{1}{2}}\,K\,.\n{3.19}
\ee
where $K\,=\,n^{\alpha}{}_{|\alpha} $ is the mean extrinsic curvature, and
$N$, $s^a $ the lapse and shift which appear in the standard ADM
form of the bulk metric
\be
ds^2\,=\,g_{ab}\,(d\xi^a \,+\,s^a\,dz)\,(d\xi^b \,+\,s^b\,dz)\,+\,
\epsilon\,N^2\,dz^2\,,\n{3.20}
\ee
so that
\be
g^{00}\,=\,\epsilon\,N^{-2}\,;\qquad g^{0a}\,=\,-\epsilon\,N^{-2}\,s^a\,;
\qquad |{}^{(n)} g|^{\frac{1}{2}}\,=\,\sqrt{-g}\,N^{-1}\,.\n{3.21}
\ee
In the anchored co-ordinates (\ref{3.16}), $n^{\mu} $ has components
\be
n_{\mu}\,=\,\epsilon\,N\,\partial_{\mu}z\,,\qquad N\,n^0\,=\,\epsilon\,,
\qquad N\,n^a\,=\,-\,s^a\,.\n{3.22}
\ee
It is now straightforward to derive (\ref{3.19}): from(\ref{3.18}), 
(\ref{3.21}) and (\ref{3.22}),
\ba\label{3.23}
\sigma^0\,-\,2\,\epsilon\,{\cal K}\,&=&\,
(-g)^{-\frac{1}{2}}\,\partial_{\mu}\lbrack (-g)\,\epsilon\, 
N^{-1}\,n^{\mu}\rbrack \,-\,
2\,\epsilon\,N^{-1}\partial_{\mu}(\sqrt{-g}\,n^{\mu} )\nonumber \\
&=&\,-\sqrt{-g}\,\epsilon\, N^{-2}\,\partial_{\mu}(N\,n^{\mu})\,=\,
\sqrt{-g}\,\epsilon\, N^{-2}\,\partial_a s^a\,.
\ea 
This extra contribution (\ref{3.23}) to the boundary action (\ref{3.17}), 
(\ref{3.19}) is ``velocity-independent'' (i.e., independent of transverse
derivatives $\partial_0 g_{\mu\nu} $), and has the same role as $f(q)$ in 
(\ref{2.17}). It could be reconverted to a volume integral and included with
${\cal L}_1 $ in the first-order bulk action in (\ref{3.7}). Its
contribution to the momentum is trivial. Moreover, it contributes 
nothing at all to the {\it jump } of momentum across an extrinsic-curvature
discontinuity, since its jump across $\Sigma $ is zero if $x^{\mu}$ and
$g_{\mu\nu}$ are continuous. This will continue to hold in the 
lightlike limit $(N\,\to\,\infty )$.

Recalling (\ref{3.5}), we conclude that
\be
\epsilon\,\lbrack\hspace{-0.1em}\vert\,{\cal K}(g,\Gamma )
\,\vert\hspace{-0.1em}\rbrack\,\,=\,C^{\mu\nu}_{\lambda}(g)\,
\lbrack\hspace{-0.1em}\vert\,\Gamma^{\lambda}_{\mu\nu}\,
\vert\hspace{-0.1em}\rbrack\,\,;
\qquad C^{\mu\nu}_{\lambda}(g)\,=\,\frac{\sqrt{-g}}{2}\,
(g^{\sigma (\mu}\delta^{\nu )}_{\lambda}\,-
\,g^{\mu\nu}\delta^{\sigma}_{\lambda})\,\partial_{\sigma}z\,,\n{3.24}
\ee 
represents precisely the jump of mean extrinsic curvature density at
an {\it arbitrary } (lightlike or non-lightlike) extrinsic-curvature 
discontinuity $\Sigma $. (For lightlike $\Sigma $, $\epsilon $ is defined by 
continuity with the timelike case.)

A surface layer can be characterized by supposing the bulk divided into
two subdomains $z<0$ and $z>0$ with edges $\Sigma_- (z=-0)$ and 
$\Sigma_+ (z=+0)$, glued together to form a common boundary $\Sigma $, 
loaded with a surface distribution of stress-energy; $\Sigma_+ $ and
$\Sigma_- $ are supposed intrinsically isometric, but 
$\partial_{\sigma}g_{\mu\nu}$ undergoes a jump at $\Sigma $ in smooth 
(e.g. skew-Gaussian) co-ordinates, i.e., $\Sigma $ is an extrinsic-curvature
discontinuity. Dynamically, the surface layer is accounted for by adding
to the Einstein-Hilbert action (\ref{3.7}) a shell contribution
equal to the sum of the actions (\ref{3.17}) for the boundaries
$\Sigma_- $ and $\Sigma_+ $ of the subdomains, taking into account
the opposite directions of their outward normals:
\be
S_{EH,shell}\,=\,\frac{1}{2\kappa}\,\int_{\Sigma}\,2\,\epsilon\,
\lbrack\hspace{-0.1em}\vert\,{\cal K}\,\vert\hspace{-0.1em}\rbrack\,
\,d^n\xi\,+\,S_{mat,shell}\,.\n{3.25}
\ee
The second (matter) term $S_{mat,shell} $ generates the surface stress-energy
density ${\cal S}^{\alpha\beta}$ via
\be
\frac{\delta S_{mat,shell}}{\delta g_{\alpha\beta}}\,\Big|_{\Sigma}\,=\,
\frac{1}{2}\,\,{\cal S}^{\alpha\beta}\,,\n{3.26}
\ee
in analogy with (\ref{3.9}). 

Variation of the total action (\ref{3.7})+(\ref{3.25}) 
then yields the jump conditions
\be
\lbrack\hspace{-0.1em}\vert\,\pi^{\alpha\beta}\,\vert\hspace{-0.1em}\rbrack\,
=\,\frac{1}{2}\,{\cal S}^{\alpha\beta}\,,\n{3.27}
\ee
in addition to the bulk field equations (\ref{3.13}).

The canonical field momentum density $\pi^{\alpha\beta}$ associated with
any boundary $\Sigma $ is defined in anchored co-ordinates (\ref{3.14}) by
\be
\pi^{\alpha\beta}\,=\,\frac{1}{2\kappa}\,\frac{\partial{\cal L}_1 }
{\partial g_{\alpha\beta ,0}}\Big|_{\Sigma}\,.\n{3.28}
\ee
It is more conveniently extracted from the Hamilton-Jacobi variational formula
\be
\delta S_{EH}\,=\,\int_{\Sigma}\,\pi^{\alpha\beta}\,\delta g_{\alpha\beta}\,
d^n \xi\,=\,\delta_g S_{bdy} (g,\Gamma )\,, \n{3.29}
\ee
modulo the bulk field equations (\ref{3.13}), and recalling (\ref{3.12}).
For the shell (\ref{3.25}), we obtain from (\ref{3.29}), (\ref{3.17}) 
and (\ref{3.19}) the jump
\be
\kappa \,\lbrack\hspace{-0.1em}\vert\,\pi^{\alpha\beta}\,
\vert\hspace{-0.1em}\rbrack\,=\,
\frac{\partial \,{\lbrack\hspace{-0.1em}\vert\,\epsilon\cal K}(g,\Gamma )
\vert\hspace{-0.1em}\rbrack\,}
{\partial g_{\alpha\beta}}\,,\n{3.30}
\ee
with $g$, $\Gamma $ treated as independent, 
in analogy with the bulk identity
\be
\sqrt{-g}\,G_{\alpha\beta}\,=\,\frac{\delta }{\delta g^{\alpha\beta}}\,
\int\,\sqrt{-g}\,R\,d^{n+1}x\,=\,\frac{\partial}{\partial g^{\alpha\beta}}\,
\left( \sqrt{-g}\,g^{\mu\nu}\,R_{\mu\nu}(\Gamma )\right)\,. \n{3.31}
\ee

Explicit evaluation of (\ref{3.30}) with the help of (\ref{3.24}) yields
\be
\lbrack\hspace{-0.1em}\vert\,\pi^{\alpha 0}\,\vert\hspace{-0.1em}\rbrack\,
=\,0\,,\n{3.32}
\ee
\ba\label{3.33}
\kappa\,
\lbrack\hspace{-0.1em}\vert\,\pi^{ab}\,\vert\hspace{-0.1em}\rbrack\,
&=&\frac{1}{4}\,\sqrt{-g}\,
\lbrack\hspace{-0.1em}\vert\,g_{cd,0}\,\vert\hspace{-0.1em}\rbrack\,
\Big\{ g^{00}\,(g^{ab}g^{cd}\,-\,g^{ac}g^{bd})\nonumber\\
&+& 2\,g^{0(a}g^{b)(c}g^{d)0}
\,-\,g^{a0}g^{b0}g^{cd}\,-\,g^{ab}g^{c0}g^{d0}\,\Big\}\,.
\ea
This is equivalent to a result (eq.(17) of Barrab\`es-Israel \cite{BIH}) 
previously derived by integration of the field equations through the
layer.

Like the Bianchi identity $G^{\alpha\beta}{}_{|\beta}\,=\,0$ for the 
bulk field equations (\ref{3.13}), the transversality condition (\ref{3.32}),
\be
\lbrack\hspace{-0.1em}\vert\,\pi^{\alpha\beta}\,\vert\hspace{-0.1em}\rbrack\,
(\partial_{\beta}z)\,=\,0\,,\n{3.34}
\ee
may be regarded as a consequence of the co-ordinate-invariance of the action. 
The boundary action (\ref{3.17}) is invariant under the infinitesimal
anchored co-ordinate transformation
\be
x^{\mu}\,\to\,\bar{x}^{\mu}\,=\,x^{\mu}\,+\,\frac{1}{2} z^2\,\eta^{\mu}(x)
\,.\n{3.35}
\ee
This gives
\be
\delta_L g_{\alpha\beta}\,=\,g_{\alpha\beta}(\bar{x})\,-\,
\bar{g}_{\alpha\beta}(\bar{x})\,\,=\,
2\,\eta_{(\alpha}\partial_{\beta)}z\,+\,O(z)\,\,,\qquad (z\,\to\,0)\,.\n{3.36}
\ee
Hence from (\ref{3.29}),
\be
\delta_{L} S_{EH}\,=\,\int_{\Sigma}\,\pi^{\alpha\beta}\,2\,\eta_{\alpha}\,
\partial_{\beta}z\,d^n \xi\,,\n{3.37}
\ee
which is required to vanish by co-ordinate invariance of the total action, 
leading to (\ref{3.33}).

For a non-lightlike layer ($g^{00}|_{\Sigma}\neq 0$), (\ref{3.33}) 
simplifies to
\be
\kappa\,\lbrack\hspace{-0.1em}\vert\,\pi^{ab}\,\vert\hspace{-0.1em}\rbrack\,
=\,\frac{1}{4}\,\sqrt{-g}\,
\lbrack\hspace{-0.1em}\vert\,g_{cd,0}\,\vert\hspace{-0.1em}\rbrack\,
g^{00}\,(\Delta^{ab}\Delta^{cd}\,-\,\Delta^{ac}\Delta^{bd})\,\n{3.38}
\ee
where $\Delta^{ab}$ projects onto $\Sigma $ and coincides with the 
inverse intrinsic metric in anchored co-ordinates
\be
\Delta^{ab}\,=\,g^{ab}\,-\,\frac{g^{a0}g^{b0}}{g^{00}}\,=\,g^{ab}\,-\,
\epsilon\,n^a n^b\,=\,{}^{(n)}g^{ab}\,.\n{3.39}
\ee
The jump conditions (\ref{3.27}) can then be reduced to their standard 
non-lightlike form \cite{MTW}
\be
-\,\frac{\epsilon }{2}\,|{}^{(n)}g|^{\frac{1}{2}}\,
\lbrack\hspace{-0.1em}\vert\,K^{ab}\,-\,{}^{(n)}g^{ab}\, K\,
\vert\hspace{-0.1em}\rbrack\,=\,\kappa\,
\lbrack\hspace{-0.1em}\vert\,\pi^{ab}\,\vert\hspace{-0.1em}\rbrack\,=\,
\frac{\kappa }{2}\,|{}^{(n)}g|^{\frac{1}{2}}\,S^{ab}\,,\n{3.40}
\ee
in terms of the jump of extrinsic curvature
\be
\lbrack\hspace{-0.1em}\vert\,K_{ab}\,\vert\hspace{-0.1em}\rbrack\,=\,
\frac{1}{2}\,\lbrack\hspace{-0.1em}\vert\,\frac{\partial g_{ab}}{\partial n}\,
\vert\hspace{-0.1em}\rbrack\,=\,\frac{1}{2}\,|g^{00}|^{\frac{1}{2}}\,
\lbrack\hspace{-0.1em}\vert\,\partial_0 g_{ab}\,\vert\hspace{-0.1em}\rbrack\,
\,,\n{3.41}
\ee
and of the surface stress-energy tensor $S^{ab} $ of the shell
\be
S^{ab}\,=\,|{}^{(n)}g|^{-\frac{1}{2}}\,{\cal S}^{ab}\,.\n{3.42} 
\ee

\section{Gauss-Bonnet Action}
\setcounter{equation}0
When the Einstein-Hilbert action (\ref{3.1}) is augmented with term 
quadratic in the curvature the simple form (\ref{3.40}) of the junction 
conditions is no longer valid. In fact, a distributional brane dynamics
is no longer even possible in general, because the bulk field equations
now involve inadmissible products of distributions in the thin-layer limit. 
The exception is the case when the quadratic terms have the Gauss-Bonnet
form. In this case the bulk field equations are quasi-linear, and a 
distributional description of thin layers remains viable. In this Section,
we shall examine how the junction conditions (\ref{3.40}) are modified.

The Gauss-Bonnet action is (see the Appendix for further details)
\be
S_{bulk}\,=\,\frac{\alpha }{2\kappa }\,\int\,{\cal L}_{GB}(g,\Gamma )\,
d^{n+1}x\, ,\n{4.1}
\ee
where $\alpha $ is the Gauss-Bonnet coupling constant, and
\be
{\cal L}_{GB}(g,\Gamma )\,=\,\sqrt{-g}\,{\cal R} \,=\,\sqrt{-g}\,\frac{1}{4}\,
\delta ^{343'4'}_{121'2'}\,g^{25}\,g^{2'5'}\,R^1{}_{534}\,R^{1'}{}_{5'3'4'}\,.
\n{4.2}
\ee

Because of the plethora of indices in such expressions, we have found
it convenient in many instances to let numerical indices $1,2,...$, 
$1',2',....$ do duty for literal indices 
$\alpha_1 ,\alpha_2 ,...$, $\alpha_{1}' ,\alpha_{2}' ,...$.
They are understood to run from $0$ to $n$, and repeated indices are to
be summed. In contrast, we reserve the index $0$ to stand just for its 
numerical self, and, as in (\ref{3.16}), $x^0 \,=\,z\,=\,0$ will represent
the boundary $\Sigma $.

There is no holonomic split, analogous to (\ref{3.4}), of ${\cal L}_{GB}$
into a first-order piece and a pure divergence. We must therefore proceed
more indirectly to find the supplementary boundary action which effectively
removes second derivatives from the GB bulk action (\ref{4.1}).

Varying the affine connection $\Gamma $ in (\ref{4.2}), and noting the 
Palatini and Bianchi identities,
$\delta R^1{}_{534}\,=\,-2\,\delta \Gamma^1_{53|4}$ and  
$\delta^{.43'4'}_{....}\,R^{1'}{}_{5'3'4'|4}\,=\,0 $,
we see that the $\Gamma $-variation of $S_{bulk}$ involves a pure divergence,
convertible to a surface integral:
\be
\frac{2\kappa }{\alpha }\,\delta_{\Gamma}S_{bulk}\,=\,\int\,\sqrt{-g}\,
\sigma^{\lambda}{}_{|\lambda}\,d^{n+1} x\,=\,
\int_{\Sigma}\,\sqrt{-g}\,\sigma^{\lambda}\,\partial_{\lambda}z\,d^n x\, ,
\n{4.3}
\ee
where
\be
\sigma^{\lambda}\,=\,-\,\delta^{3\lambda 3'4'}_{121'2'}\,g^{25}\,
R^{1'2'}{}{}_{3'4'}\,\delta\Gamma^1_{53}\,. \n{4.4}
\ee
This is in complete analogy with (\ref{3.11}) in the Einstein-Hilbert case,
and it means that the bulk field equations
\be
\frac{1}{2\kappa}\,
\frac{\partial {\cal L}(g,\Gamma )}{\partial g^{\mu\nu }}\,+\,
\frac{1}{2}\,\frac{\delta S_{mat}}{\delta g^{\mu\nu }}\,=\,0\,,\n{4.5}
\ee
with ${\cal L}\,=\,{\cal L}_{EH}\,+\,\alpha\,{\cal L}_{GB}$,
are obtainable by simply differentiating the bulk Lagrangians (\ref{4.2}) and
(\ref{3.2}) partially with respect to $g$.

When the bulk field equations (\ref{4.5}) are satisfied, the boundary 
term (\ref{4.3}) gives the total variation of the bulk action. This involves 
$\delta \Gamma $, from which the variations $\delta g_{\alpha\beta ,0}$ of 
transverse derivatives --i.e., of extrinsic curvature in the non-lightlike 
case -- must be removed by compensating variation of a suitable boundary
action.

To isolate these extrinsic curvature variations, we assume for simplicity
that $\Sigma $ is non-lightlike and introduce Gaussian co-ordinates 
based on $\Sigma $ as in (\ref{3.20}). Then
\be
{}^{(n+1)}\Gamma^c_{ab}\,=\,{}^{(n)}\Gamma^c_{ab}\,\qquad 
{}^{(n+1)}\Gamma^0_{ab}\,=\,\epsilon\,K_{ab}\,,\qquad 
{}^{(n+1)}\Gamma^a_{0b}\,=\,K^a{}_b\,,\n{4.6}
\ee
where the extrinsic curvature $K_{ab}\,=\,\frac{1}{2}\,\partial_0 g_{ab}$,
and Latin indices run from $1$ to $n$. We note also the Gauss-Codazzi relations
\be
{}^{(n+1)}R_{abcd}\,=\,{}^{(n)}R_{abcd}\,-\,2\,\epsilon\,
K_{a[c}K_{d]b}\,,\n{4.7}
\ee
\be
{}^{(n+1)}R_{\mu bcd}\,n^{\mu}\,=\,2\,K_{b[c;d]}\,,\n{4.8}
\ee
where $;$ represents the covariant derivative associated with the 
$n$-dimensional metric $g_{ab}$.
Retaining only the variations $\delta K_{ab}$ in (\ref{4.4}) -- i.e., assuming
$\delta g_{\alpha\beta}|_{\Sigma}\,=\,0$ -- we find
\be
\sigma^0\,=\,-\,\delta^{303'4'}_{101'2'}\,R^{1'2'}{}{}_{3'4'}\,2\,\epsilon\,
\delta K^1{}_3\,, \n{4.9}
\ee
so that (\ref{4.3}) becomes
\be
\frac{2\kappa }{\alpha }\,\delta_K \,S_{bulk}\,=\,
-\,2\,\epsilon \,\delta^{033'4'}_{011'2'}\,\int_{\Sigma}\,\left( 
{}^{(n)}R^{1'2'}{}{}_{3'4'}\,-\,
2\,\epsilon\,K^{1'}{}_{3'}\,K^{2'}{}_{4'}\right)\,
\delta K^1{}_3\,\sqrt{-g}\,d^n x\,,\n{4.10}
\ee
where we have made use of (\ref{4.7}).

The boundary action $S_{bdy}$ must be chosen so that its $K$-variation cancels
(\ref{4.10}):
\be
\delta_K (S_{bulk}\,+\,S_{bdy}\,)\,=\,0\,,\n{4.11}
\ee
Since the intrinsic Riemann tensor ${}^{(n)}R$ is independent of $K_{ab}$,
the choice
\be
\frac{2\kappa }{\alpha }\,S_{bdy}\,=\,2\,\epsilon \,\delta^{033'4'}_{011'2'}\,
\int_{\Sigma}\,\left( 
{}^{(n)}R^{1'2'}{}{}_{3'4'}\,-\,
\frac{2\epsilon}{3}\,K^{1'}{}_{3'}\,K^{2'}{}_{4'}\right)\,K^1_3\,
\sqrt{-g}\,d^n x\,,\n{4.12}
\ee
meets this requirement, a result originally due to Myers \cite{My}.

We are now ready to derive the Gauss-Bonnet field momentum $\pi^{ab}$ 
associated with $\Sigma $ from the Hamilton-Jacobi equation
\be
\delta (S_{bulk}\,+\,S_{bdy}\,)\,=\,\int_{\Sigma}\,\pi^{ab}\,\delta g_{ab}\,
d\Sigma \,,\n{4.13}
\ee
modulo the field equations (\ref{4.5}).

Evaluation of the left-hand side of (\ref{4.13}) is simplified by noting
that there is no $K$-contribution because of (\ref{4.11}), 
and none from the metric 
factors in (\ref{4.2}) because of (\ref{4.5}). So $\delta S_{bulk}\,=\,
\delta_{\Gamma}S_{bulk}$ is again given by (\ref{4.3}) and (\ref{4.4}),
with $\delta \Gamma $ now effectively determined solely by the metric 
variation $\delta g_{ab}$, with $K_{ab}$ fixed. We re-express (\ref{4.4}) as
\be
\sigma^0\,=\,-2\,\delta^{303'4'}_{121'0}\,g^{25}\,
R^{1'0}{}{}_{3'4'}\,\delta\Gamma^1_{53}\,
=\,4\,\epsilon\,\delta^{03'34}_{011'2'}\,K^1{}_{3;4'}\,g^{2'5'}\,
\delta\Gamma^{1'}_{5'3'}\, ,\n{4.14}
\ee
where we have interchanged primed and unprimed indices in the second 
line and made use of the Gauss-Codazzi relation(\ref{4.8}).

Turning now to $\delta S_{bdy}$ in (\ref{4.13}), one piece of this arises 
from variation of the intrinsic curvature ${}^{(n)}R $ in (\ref{4.12}):
\be
\frac{2\kappa }{\alpha}\,\delta S_{bdy}\,\to\,2\,\epsilon\,
\delta^{033'4'}_{011'2'}\,\int_{\Sigma}\,g^{2'5'}\,
\delta{}^{(n)}R^{1'}{}_{5'3'4'}\,K^1{}_3\,|{}^{(n)}g|^{\frac{1}{2}}\,d^n\,x\,.
\n{4.15}
\ee
Applying the intrinsic Palatini identity
$ \delta{}^{(n)}R^{1'}{}_{5'3'4'}\,=\,-2\,\delta \Gamma^{1'}_{5'[3';4']}$, 
one sees that the integrands (\ref{4.15}) and (\ref{4.14}) add up to an 
intrinsic divergence, which may be discarded.

All that remains to account for is the metric variations arising from 
$\sqrt{-g}$, the raised indices ($2'$ on ${}^{(n)}R^{..}{}{}_{..}$, 
$1'2'1$ on the $K$-factors) in (\ref{4.12}), and from
$\delta \Gamma^1_{03}\, =\,\delta g^{1b}\, K_{b3}$
in the variation of the bulk action (4.3), (4.4). 
The result of a straightforward calculation is
\ba\label{4.16}
\frac{2\kappa}{\alpha}\,\pi_{ab}\,&=& |{}^{(n)}g|^{\frac{1}{2}}\,
\Big\{\, 6\,K_{a[m}\,K^m_n\,K^n_{b]}\,+\, 6\,K_{b[m}\,K^m_n\,K^n_{a]}\,
\nonumber\\
&-&\,4g_{ab}\,K^l_{[l}\,K^m_m\,K^n_{n]}\,
+\, 4\epsilon\,K^{cd}\,\,{}^{*(n)}R^*{}{}_{acbd}\,\Big\}\,.
\ea
where ${}^{*(n)}R^*{}{}_{ac}{}{}^{bd}$ -- see the equation (\ref{A.6'}) of
the Appendix for its definition -- 
is the left and right dual of
the intrinsic curvature tensor of $\Sigma $. 
This is equivalent to results previously obtained by Davis and others 
\cite{Dav}.

Following the argument leading to (\ref{3.27}), we conclude that the
dynamics of a {\it non-lightlike } shell in Einstein-Gauss-Bonnet theory,
with bulk action
\be
S_{bulk}\,=\,\frac{1}{2\kappa}\,\int\,\sqrt{-g}\,(R\,+\,\alpha {\cal R}\,)\,
d^{n+1}x\,+\,S_{mat}\,,\n{4.17}
\ee
is governed by the junction conditions
\be
\lbrack\hspace{-0.1em}\vert\,\pi^{ab}\,\vert\hspace{-0.1em}\rbrack\,=\,
\frac{|{}^{(n)}g|^{\frac{1}{2}} }{2}\,S^{ab}\,;\qquad 
\pi^{ab}\,=\,\pi^{ab}_{EH}\,+\,\pi^{ab}_{GB}\,,\n{4.18}
\ee
where $S^{ab} $ is the surface stress-energy tensor of the shell, and is
defined by (\ref{3.26}) and (\ref{3.42}). 
The momenta $\pi^{ab}_{EH}$ and $\pi^{ab}_{GB}$ are given by 
the first of (\ref{3.40}) and
(\ref{4.16}) respectively, and jump together with the extrinsic curvature
across the shell. The action due to the shell augments (\ref{4.17}) with 
a surface term --cf (\ref{3.25}) and (\ref{4.12})-- and is equal to 
\be
S_{shell}\,=\,\frac{1}{2\kappa}\,\int_{\Sigma}\,
\left( 
\lbrack\hspace{-0.1em}\vert\,B_{EH}\,\vert\hspace{-0.1em}\rbrack\,+\,\alpha\,
\lbrack\hspace{-0.1em}\vert\,B_{GB}\,\vert\hspace{-0.1em}\rbrack\,
\right)\,d\Sigma\,+\,
S_{mat,shell}\,,\n{4.19}
\ee
where
\be
 B_{EH}\,=\,2\,\epsilon\,K\,\n{4.20}
\ee
\be
B_{GB}\,=\,2\,\epsilon\,\delta^{cdf}_{abe}\,\left( 
{}^{(n)}R^{ab}{}{}_{cd}\,-\,\frac{2\epsilon}{3}\,K^a{}_c\,K^b{}_d\right) 
\,K^e_f\,.\n{4.21}
\ee

These results hold for non-light-like shells, for which extrinsic curvature
is well-defined. We now add some remarks on the lightlike case. Since
this is a special limit of the timelike case, one might at first sight
expect the resulting junction conditions to be simpler. Actually, however,
this is far from being the case. Lightlike discontinuities propagate
along characteristics of the field equations. It is a nontrivial matter
to disentangle the lightlike discontinuities due to the shell from the
accompanying gravitational shock waves. For pure Einstein theory
this is still quite manageable, and we have presented the results in Sec. 3.
But we have not yet succeeded in reducing the lightlike junction conditions for
Einstein-Gauss-Bonnet theory to a form that we consider worth publishing. 
The nature of the characteristics themselves is made more complicated by
the fact that the field equations are now quasiliniear only in the
broad sense (linear in second derivatives but with coefficients depending
on first derivatives).

It is, however, straightforward to obtain the Gauss-Bonnet boundary and shell
actions in the lightlike case. Under an arbitrary metric variation, the
total variation of the bulk action, is still given (modulo the bulk field
equations) by (\ref{4.3}) and (\ref{4.4}), which we can express in the form
\be
\sigma^0\,=\,-4\,{}^{*}R^{*\,3012}\,\delta\Gamma_{1,23}\,,\n{4.22}
\ee
where we have introduced the left and right dual of the 
curvature tensor, defined by (\ref{A.3}) and (\ref{A.6}) in the Appendix. 
In (\ref{4.22}) we must separate out (and neutralize 
with a boundary action) the contribution of variations 
$\delta g_{\alpha\beta ,0}$ of transverse derivatives:
\be
\delta\Gamma_{[1,2]3}\,=\,-\,\partial_{[1}\delta g_{2]3}\,\to\,
-\,(\partial_{[1} z)\,\delta g_{2]3,0}\,.\n{4.23}
\ee
Hence the boundary action must have the compensating variation
\be
\frac{2\kappa}{\alpha}\,\delta S_{bdy}\,=\,
4\,\int\,{}^{*}R^{*\,0203}\,\delta g_{23,0}\,\sqrt{-g}\,d^n x\,.\n{4.24}
\ee
Now, when the surfaces, $x^0\,=\,const.$ are lightlike, the components
${}^{*}R^{*\,0203}$ do not contain second transverse derivatives
$g_{\alpha\beta ,00}$ and they are linear in first derivatives, i.e.,
\be
{}^{*}R^{*\,0a0b}\,=\,K^{ab}_{(0)}\,+\,K^{abcd}_{(1)}\,g_{cd,0}\,;\qquad
K^{abcd}_{(1)}\,=\,K^{cdab}_{(1)}\,,\n{4.25}
\ee
where $K_{(0)}$, $K_{(1)}$ are independent of $g_{ab,0}$ ($K_{(1)}$ has a 
fairly complicated linear dependence, not reproduced here, on the ``nominal
extrinsic curvature'' ${\cal K}_{ab}\,=\,-\,\Gamma^0_{ab}$, 
which depends only on the intrinsic geometry for a lightlike 
$\Sigma $ ). From (\ref{4.24})
and (\ref{4.25}) we infer
\be
\frac{2\kappa}{\alpha}\,S_{bdy}\,=\,4\,\int\,\left( K_{(0)}^{ab}\,+\,
\frac{1}{2}\,K_{(1)}^{abcd}\,
g_{cd,0} \right)\,g_{ab,0}\,\sqrt{-g}\,d^n x\,,\n{4.26}
\ee
as the form of the Gauss-Bonnet boundary action.

\section{Concluding Remarks}

We have given an elementary, self-contained derivation of the action
(useful for calculation of quantum tunneling amplitudes) and dynamical
equations (i.e. junction conditions) for thin shells and branes in 
Einstein-Gauss-Bonnet theory. Our exposition has attempted
to integrate as far as possible the treatment of timelike, spacelike 
and lightlike layers. For lightlike shells, the dynamics is 
complicated (especially in the Gauss-Bonnet case) by the fact that 
gravitational shock waves will in general accompany the shell.
However, this problem should be ameliorated in situations of
high symmetry, and this is currently under investigation.

\appendix
\setcounter{equation}{0}
\section{Notations for GB}\indent

In 1932, Lanczos \cite{Lan} noted that the Lagrangian
\be
{\cal R}\,=\,\frac{1}{4}\,\delta^{343'4'}_{121'2'}\,R^{12}{}{}_{34}\,
R^{1'2'}{}{}_{3'4'}\,,\n{A.1}
\ee
leads, like the Einstein-Hilbert Lagrangian
\be
R\,=\,\frac{1}{2}\,\delta^{34}_{12}\,R^{12}{}{}_{34}\,,\n{A.2}
\ee
to field equations which involve no higher than second derivatives of the
matric. $R$ and ${\cal R}$ are the first two members of a family 
of Lagrangians having the 
same property found by Lovelock \cite{Lov}. The $n^{th}$ member involves 
a product of $n$ curvature factors formed by an obvious generalization of
(\ref{A.1}), (\ref{A.2}). It has the ``Gauss-Bonnet'' property of being a 
pure divergence in a space of dimension $2n$, and it vanishes identically in
spaces of lower dimension. Properties of the Lovelock family are reviewed by 
Meissner and Olechowski \cite{Zw} and Deruelle and Madore \cite{Zw}.

By defining the left and right dual of the curvature tensor
\be
{}^{*}R^*{}{}_{12}{}{}^{34}\,=\,\frac{1}{4}\,\delta^{343'4'}_{121'2'}\,
R^{1'2'}{}{}_{3'4'}\,,\n{A.3}
\ee  
and using the identity
\ba\label{A.4}
\delta^{343'4'}_{121'2'}\,&=&\,4!\,
\delta^{[3}_1\,\delta^4_2\,\delta^{3'}_{1'}\,\delta^{4']}_{2'}\nonumber\\
&=&\delta^{34}_{12}\,\delta^{3'4'}_{1'2'}\,+\,
\delta^{3'4'}_{12}\,\delta^{34}_{1'2'}\,-\,
2\,\delta^{3[3'}_{12}\,\delta^{|4|4']}_{1'2'}\,\,-\,
2\,\delta^{4[4'}_{12}\,\delta^{|3|3']}_{1'2'}\,,
\ea
where $\delta^{34}_{12}\,=\,\delta^{3}_{1}\,\delta^{4}_{2}\,
-\,\delta^{4}_{1}\,\delta^{3}_{2}\,$, (\ref{A.1}) and (\ref{A.3}) can be 
recast as
\be
{\cal R} \,=\,R^{12}{}{}_{34}\,{}^{*}R^*{}{}_{12}{}{}^{34}\,=\,
(R_{1234})^2\,-\,4\,(R_{12})^2\,+\,R^2\,,\n{A.5}
\ee
(which was the form originally given by Lanczos). 
The tensor ${}^{*}R^*{}{}_{12}{}{}^{34}$ generalizes to spacetime with
dimension higher than four the left and right dual of the curvature tensor
of general relativity. It is equal to
\be
{}^{*}R^*{}{}_{12}{}{}^{34}\,=\,R_{12}{}{}^{34}\,-\,4\,
\delta^{[3}_{[1}\,R^{4]}{}{}_{2]}\,+\frac{1}{2}\,\delta^{34}_{12}\,R\,
=\,R_{12}{}{}^{34}\,-\,4\,
\delta^{[3}_{[1}\,G^{4]}{}{}_{2]}\,-\frac{1}{2}\,\delta^{34}_{12}\,R\,
.\n{A.6}
\ee
A similar quantity ${}^{*(n)}R^*{}{}_{ab}{}{}^{cd}$ can be defined for
the intrinsic curvature tensor
\be
{}^{*(n)}R^*{}{}_{ab}{}{}^{cd}\,=\,\frac{1}{4}\,\delta^{cdc'd'}_{aba'b'}\,\,
{}^{(n)}R^{a'b'}{}{}_{c'd'}\,=\,{}^{(n)}R_{ab}{}{}^{cd}\,-\,4\,
\delta^{[c}_{[a}\,\,{}^{(n)}R^{d]}{}{}_{b]}\,+
\frac{1}{2}\,\delta^{cd}_{ab}\,\,{}^{(n)}R\,
,\n{A.6'}
\ee   
where the latin indices run from $1$ to $n$.

Because, as noted in Sec. 4, the $\Gamma $-variations of $R$ and ${\cal R}$ 
are pure divergences, the bulk field equations
\be
G_{\mu\nu}\,+\,\alpha\,{\cal G}_{\mu\nu}\,=\,\kappa\,T_{\mu\nu}\,,\n{A.7}
\ee
can be obtained from the bulk action
\be
S_{bulk}\,=\,\frac{1}{2\kappa}\,\int \sqrt{-g}\,
(\,R\,+\,\alpha\,{\cal R}\,)\,+\,
S_{mat}\,,\n{A.8}
\ee
simply by partially differentiating the Lagrangian, 
holding $R^1{}{}_{234}(\Gamma )$ fixed:
\be
{\cal G}_{\mu\nu}\,=\,\frac{1}{\sqrt{-g}}\,
\frac{\partial }{\partial g^{\mu\nu}}(\sqrt{-g}\,{\cal R}\,)\,
=\,2\,{\cal R}_{\mu\nu}\,-\,
\frac{1}{2}\,g_{\mu\nu}\,{\cal R}\,,\n{A.9}
\ee
where
\be
{\cal R}\,=\,g^{\mu\nu}\,{\cal R}_{\mu\nu}\,;\qquad
{\cal R}_{\mu\nu}\,=\,\frac{1}{2}\,g^{\alpha\beta}\,
(\,{\cal R}_{\alpha\mu\beta\nu}\,+\,{\cal R}_{\alpha\nu\beta\mu})\,,\n{A.10}
\ee
\be
{\cal R}_{\alpha\beta\mu\nu}\,=\,\frac{1}{4}\,\delta^{1'2'34}_{1\alpha 3'\beta}
\,R^1{}{}_{\mu 34}\,R^{3'}{}{}_{\nu 1'2'}\,.\n{A.11}
\ee
(\ref{A.9}) confirms that the field equations follow the action
in containing no higher than second derivatives of the metric.
Noether's theorem and co-ordinate-invariance of $R$ and ${\cal R}$ imply
the contracted Bianchi identities
\be
G^{\alpha\beta}{}{}_{|\beta}\,=\,{\cal G}^{\alpha\beta}{}{}_{|\beta}\,=\,0
\,,\n{A.12}
\ee
which ensure compatibility of (\ref{A.7}) with the conservation law
$T^{\alpha\beta}{}{}_{|\beta}\,=\,0 $.
All the above considerations extend straightforwardly to higher members
of the Lovelock family.

\noindent
\section*{Acknowledgment}\noindent
The present work was partly supported by the NATO Collaborative
Linkage Grant (979723), by the Centre National de la
Recherche Scientifique (France), and by NSERC of Canada.

\end{document}